\documentclass[aps,prb,groupedaddress,twocolumn,showpacs]{revtex4-1}

\usepackage{relsize}
\usepackage{amsmath,amssymb,bbm,bm}
\usepackage{graphicx}

\newcommand{\vect}[1] { {\bm{#1}} }
\newcommand{\calF}{\mathcal{F}}
\newcommand{\calH}{\mathcal{H}}
\newcommand{\calT}{\mathcal{T}}
\newcommand{\calU}{\mathcal{U}}

\newcommand{\sig}{\vect \sigma}
\newcommand{\xproduct}{\!\times\!}
\newcommand{\sproduct}{\!\cdot\!}
\newcommand{\ez}{\vect {\hat e_z} }

\newcommand{\ket}[1]{| {#1} \rangle}
\newcommand{\bra}[1]{\langle {#1} |}
\newcommand{\upd}{ {\rm d} }

\newcommand{\Lidx}{{\raisebox{-2pt}{\tiny L}}}
\newcommand{\Ridx}{{\raisebox{-2pt}{\tiny R}}}
\newcommand{\LRidx}{{\raisebox{-2pt}{\tiny L/R}}}
\newcommand{\RLidx}{{\raisebox{-2pt}{\tiny R/L}}}
\newcommand{\Lid}{{\rm \scriptscriptstyle L}}
\newcommand{\Rid}{{\rm \scriptscriptstyle R}}
\newcommand{\LRid}{{\rm \scriptscriptstyle L/R}}
\newcommand{\RLid}{{\rm \scriptscriptstyle R/L}}

\newcommand{\UP} { \uparrow }
\newcommand{\DOWN} { \downarrow }
\newcommand{\UPDOWN} { {\uparrow,\downarrow} }

\newcommand{\eF}  { E_{\rm F} }

\begin{document}

\title{Reflectionless Transport of Surface Dirac Fermions on Topological Insulators with Induced Ferromagnetic Domain Walls}
\author{Christian Wickles}
\email[]{christian.wickles@uni-konstanz.de}
\affiliation{Universit\"at Konstanz, Fachbereich Physik, 78457
  Konstanz, Germany}
\author{Wolfgang Belzig}
\email[]{wolfgang.belzig@uni-konstanz.de}
\affiliation{Universit\"at Konstanz, Fachbereich Physik, 78457
  Konstanz, Germany}

\date{\today}

\begin{abstract}
   The properties of surface Dirac Fermions on a 3D topological insulator in proximity to a magnetic insulator with spatially textured magnetization are considered. We present an exact analytical treatment of the domain wall resistance and the spectrum for an extended generic domain wall with in-plane and out-of-plane magnetizations. In the latter case, we find oscillations in the domain wall resistance as a function of the wall width and for certain widths a complete absence of reflections for all incoming momenta. The surprising occurrence of oscillations and the reflectionless potentials can be related to a supersymmetry of the surface Dirac Hamiltonian combined with the domain wall profile.
\end{abstract}



\maketitle

\section{Introduction and Model}
Three dimensional strong topological insulators (3D-STI) are a recently discovered class of materials, with a prominent example being Bi$_2$Se$_3$ \cite{Hsieh2008}. Contrary to ordinary insulators, they exhibit topologically protected surface states with characteristic spin-momentum coupling, a result of strong spin-orbit interactions \cite{Zhang2009}. To exploit the full potential of these materials and for possible applications, a combination of the surface states with more conventional materials like ferromagnets or superconductors in proximity structure are desired. E.g., induced superconductivity is predicted to give rise to Majorana Fermions \cite{FuKane2009,Akhmerov2009,Tanaka2009,Linder2010a} and induced magnetization textures exhibit a quantized magneto-electric effect \cite{QiXiaoTopologicalFieldTheory2008,Qi2008NAT}. Ref. \onlinecite{QiZhangRMP2011} is a recent review covering topological states of matter.

Proximity induced ferromagnetism, where the order parameter can in general be inhomogeneous and time-dependent gives rise to phenomena interesting for spintronics and magnetotransport \cite{Mondal2010,Garate2010,Yokoyama2011,Tserkovnyak2012}. Due to the spin-momentum locking, electrical current flow leads to a significant contribution to the spin-torque acting on the magnetization dynamics \cite{Garate2010,Yokoyama2011,Tserkovnyak2012}. Magnetically doped 3D-STI could be used as a condensed matter realization of axion-electrodynamics \cite{Li2010}. The transport of Dirac-Fermions through DWs has been also studied in Graphene \cite{Yokoyama2011Graphene}.

\begin{figure}[t]
  \centering
  \includegraphics[width=7cm]{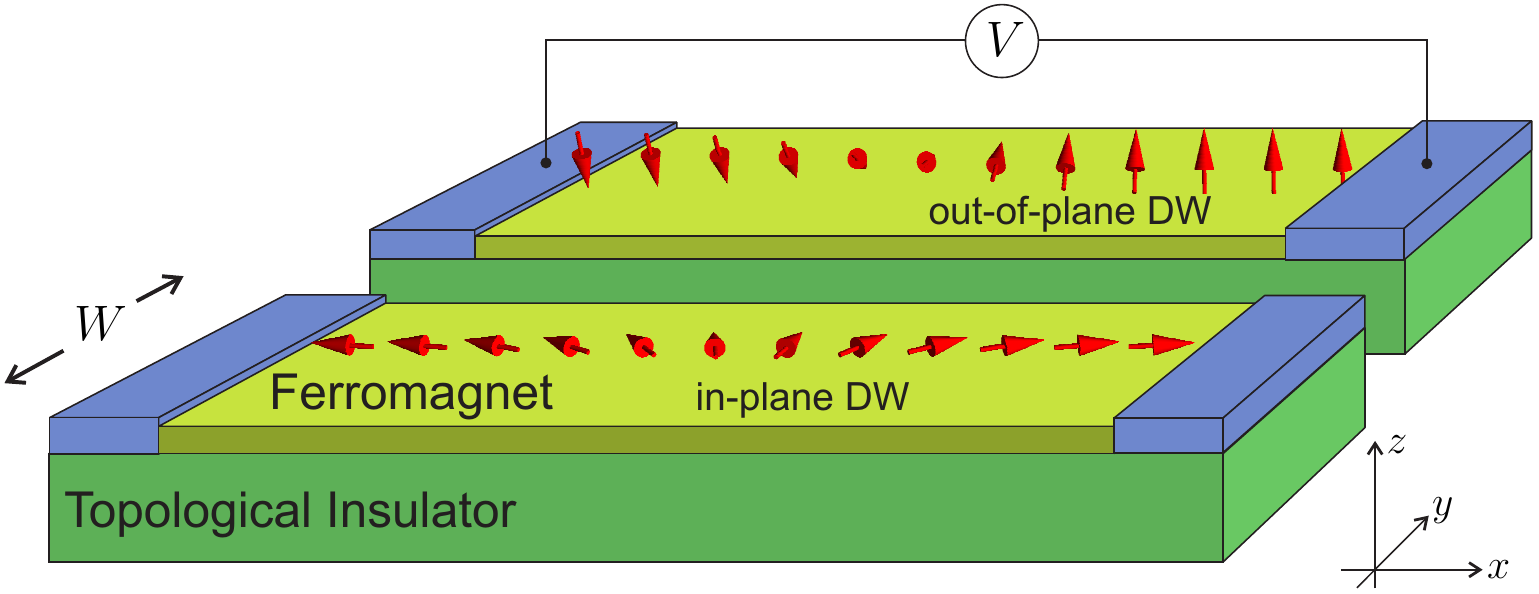}
  \caption{\label{fig:SketchOfSystem} The experimental setup for studying ballistic transport through a domain wall. Two leads with external voltage bias $V$ are attached to the structure of topological insulator coated with a isolating ferromagnetic layer that contains an in-plane (IP) tail-to-tail DW or out-of-plane (OOP) DW. }
\end{figure}

In this Letter, we consider a transport configuration in which a single static domain wall (DW) is located between two contacts as illustrated in Fig.~\ref{fig:SketchOfSystem}. We calculate the ballistic conductance for in-plane (IP) and out-of-plane (OOP) wall configurations. As our main result, in the OOP case, we find characteristic oscillations in the conductance as a function of the wall width and/or the strength of the induced exchange potential. Such a signature for the surface states in transport experiments could be visible, even if residual bulk carrier density transport is present \cite{Ren2011}.
Interestingly, we find that the DW constitutes a reflectionless potential for certain wall widths.


The general setup we study is illustrated in Fig.~\ref{fig:SketchOfSystem} and described by the effective Hamiltonian
for the two-dimensional surface electrons \begin{align}
H = i\hbar v \sig \sproduct (\ez \xproduct \vect \nabla) - M \vect m(\vect r) \sproduct \sig \ .
\end{align}
Here, the first term is the dispersion of the surface Dirac states \cite{Hsieh2009} and the second term is the proximity induced exchange coupling to the magnetization profile $\vect m(\vect r)$ with constant magnitude $M$. Such a magnetization texture will occur naturally in films of magnetic materials, which we assume to be placed on top the topological insulator. The structure can in some limits be manipulated by an external magnetic field. We stress that the induced magnetization affects only the surface states and leaves the bulk conductivity of the 3D-STI unaffected - thus opening a possible path for disentangling the surface and bulk contributions to the conductivity.

\section{The in-Plane Configuration}  The lower domain wall sketched in Fig. \ref{fig:SketchOfSystem} has the explicit form $\vect m(\vect r) = \left( \cos\vartheta(x) , \sin\vartheta(x) , 0 \right)$. We assume the angle $\vartheta(x)$ has the analytical form $\cos\vartheta(x) = \tanh\left(x/w\right)$. This shape can be obtained within a mean-field model with ferromagnetic exchange constant $J$ and anisotropy constant $K$, so that the length of the domain wall becomes $w = \sqrt{J/K}$ \cite{Tatara2004}.
Our problem is effectively one-dimensional and $k_y$ is a good quantum number. We make the ansatz for the wave-function
$\ket{\Psi(x,y)} = e^{i \frac {M w} {\hbar v} \vartheta(x)} e^{i y k_y} \ket {\psi(x)}$, by which the $m_y$-component is eliminated by the gauge factor, and end up with
\begin{align}
\label{eq:HamiltonianDWIPConfiguration}
H_{\rm IP} = i\hbar v\sigma_y \partial_x + (\hbar v k_y - M \tanh(x/w)) \sigma_x \ .
\end{align}
From now on, we choose units such that $\hbar v = 1$ and restore them only in the final results.

\begin{figure}[t]
 \centering\includegraphics[width=8.6cm]{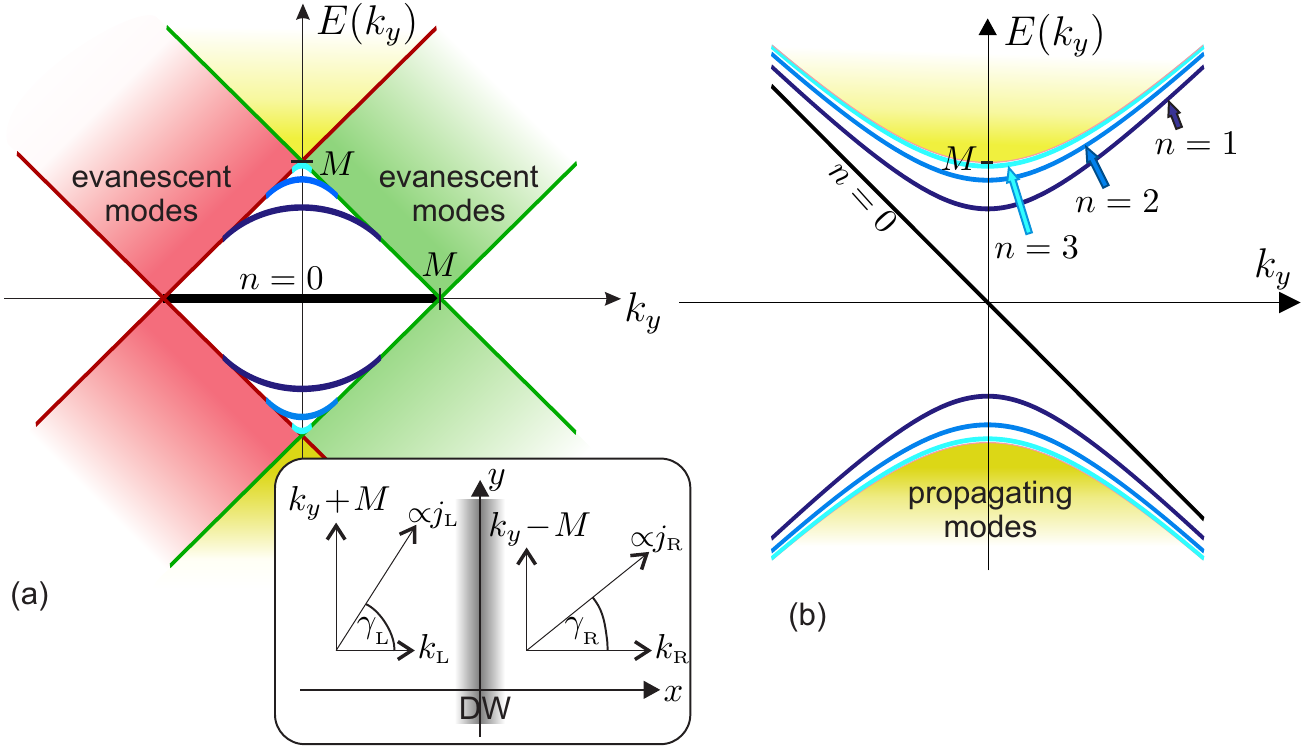}
 \caption{\label{fig:SpectrumDW} (Color online) Energy spectrum of the system with in-plane (a) and out-of-plane DW (b) including the bound states for $Mw=3.9\hbar v$. The spectrum is symmetric with respect to positive and negative energies in both cases. The continuum of scattering states (yellow area) has an energy gap $2M$ in both cases.
Continuum states that exist only on either side of the DW (red and green areas in (a)) are decaying into the potential step posed by the DW. The inset illustrates various momenta relevant for the scattering states.}
\end{figure}

In the homogeneous parts, we observe that the magnetization shifts the Dirac cone along the $k_y$-direction, which is also evident in the spectrum of this system shown in Figure \ref{fig:SpectrumDW}a. The red and green cones correspond to the dispersions far away from the DW, so that the wave vectors in transport direction for the left (L) and the right (R) side obey
$ k_\LRid^2 = E^2 - (k_y \pm  M)^2 $
when we consider an eigenstate with energy $E$. The current operator is $\vect j = (j_{x},j_{y})=v(-\sigma_{y},\sigma_{x})$. We find for incoming plane-wave states that the current points along $(k_\LRid,k_y \pm M)$ corresponding to an angle
$\gamma_{\LRidx}=\arctan((k_y \pm M)/k_\LRid)$ (see inset Fig. \ref{fig:SpectrumDW}a).

We can solve the full eigenproblem $H_{\rm IP} \ket{\psi} = E \ket{\psi}$ analytically by first transforming the two coupled first-order differential equations into two
decoupled 2nd-order differential equations for $\begin{pmatrix} \varphi_\UP,  & \varphi_\DOWN \end{pmatrix} = \ket\psi$.
In order to bring these equations into hypergeometric form, we perform the substitution $z \equiv \frac12 \left(1 - \tanh \frac xw \right)$ and subsequently make the ansatz $\varphi_\UP(z) = (1-z)^{-\frac i2 k_\Lid w} z^{-\frac i2 k_\Rid w} \ \phi(z)$ and find that
\begin{align}
\left\{z(1-z) \partial_z^2 + (\gamma - (\alpha+\beta+1)z) \partial_z - \alpha \beta \right\} \phi = 0
\end{align}
with the coefficients $\alpha = 1 + (M-ik_0) w$, $\beta = -(M+ik_0) w$ and $\gamma = 1 - i k_\Rid w$, using $k_0 = \frac12(k_\Lid + k_\Rid)$.
This equation is known \cite{Gradshteyn2007} and we directly write the solution
\begin{align}
\label{eq:HypergeometricSolutionPhiUp}
\varphi_\UP(z) = \frac{e^{-\frac i2 (\gamma_\Ridx - \frac\pi2)}}{\sqrt{k_\Rid w}} (1-z)^{-\frac i2 k_\Lid w} z^{-\frac i2 k_\Rid w} \, _2F_1(\alpha, \beta, \gamma, z) \ ,
\end{align}
where $_2F_1$ is the hypergeometric function (see appendix A for a short reference) and the prefactor is chosen such that it has preferable properties with respect to symmetry operations. More specifically, it is convenient to choose our solutions such that they are eigenstates of the operator $\calT_M \calT_{k_y} \sigma_y$ (we define symmetry operations as $\calT_M : M \rightarrow -M$ and similarly for $k_{y}$ and $x$) and as such reflect the symmetry in our system, $\left[ \calT_M \calT_{k_y} \sigma_y, H_{\rm IP} \right] = 0$. Since $(\calT_M \calT_{k_y} \sigma_y)^2 = 1$, its eigenvalues are $\pm1$, from which we can infer that the second component is simply given as $\varphi_\DOWN(z) = \mp i \calT_M \calT_{k_y}\varphi_\UP(z) $ (see below for the sign choice).

Eq.~(\ref{eq:HypergeometricSolutionPhiUp}) exhibits the different types of solutions illustrated in Fig. \ref{fig:SpectrumDW}a, depending on whether $k_\LRid$ are real or imaginary. In particular, the bound state spectrum follows for $k_\Lid$ and $k_\Rid$ being imaginary and we straightforwardly (see Appendix B for more details) find  $2\lfloor \Delta \rfloor + 1$ bound states at energy $E_{n,k_y} = \pm\frac{\hbar v}{w}\sqrt{\left(2 \Delta - n\right) n} \sqrt{1- w^2 k_y^2 /(\Delta - n)^2}$, where we defined $\Delta \equiv \frac w {l_M} = \frac {Mw}{\hbar v}$ with the magnetic length $l_M$. The evanescent modes are obtained when exactly one of $k_\Lid$ and $k_\Rid$ is imaginary.

The scattering solutions are described by real $k_\LRid$ and are pairwise degenerate. We can use parity symmetry in the $x$-$y$-plane, i.e. $\left[ H_{\rm IP} , \calT_{k_y} \calT_x \sigma_z \right] = 0$, in order to obtain the second orthogonal solution, $\ket{\psi_{-k_\Rid,k_y}^{\rm (IP)}} = -i
\calT_{k_y} \calT_x \sigma_z \ket{\psi_{k_\Lid,k_y}^{\rm (IP)}}$.
Here, $\ket{\psi_{k_\Lid,k_y}^{\rm (IP)}}$ describes an incoming wave from the left that is partially reflected and transmitted to the right side, and likewise, $\ket{\psi_{-k_\Rid,k_y}^{\rm (IP)}}$ describes an incoming wave from the right side. We note that $\calT_M \calT_{k_y} \sigma_y \ket{\psi_{\pm k_\LRid,k_y}^{\rm (IP)}} = \mp \ket{\psi_{\pm k_\LRid,k_y}^{\rm (IP)}}$.
From the asymptotic expansion of Eq.~(\ref{eq:HypergeometricSolutionPhiUp}), we can extract the transmission and reflection amplitudes and find for the tanh DW-profile the transmission probability (see Appendix B)
\begin{align}
\label{eq:TransmissionProbabilityDWIP}
	T_{\rm IP}(E, k_y) = \frac{ \sinh(\pi k_\Lid w) \sinh(\pi k_\Rid w)}{\sin^2 (\pi \Delta) + \sinh^2\!\left( \frac{\pi}{2} (k_\Lid + k_\Rid) w \right) } \ ,
\end{align}
and the reflection probability $R_{\rm IP} = 1 - T_{\rm IP}$. We remark that the transmission through evanescent or bound states is not described by Eq. (\ref{eq:TransmissionProbabilityDWIP}), i.e. we ignore any effects due to finite size geometry. This requires that the leads are sufficiently far away from the DW so that transport via the few relevant evanescent modes can be neglected over the contribution from the propagating modes. However, when the Fermi level $E_{\rm F}$ approaches $M$ and eventually reaches the crossing point of the left and right Dirac cones (crossing of red and green line in Fig. \ref{fig:SpectrumDW}a), the number of states available for transport drastically reduces and one reaches the point of minimal conductance. At this point, transport is due to few evanescent modes \cite{Katsnelson2006,Tworzydo2006}, which is beyond the present study.

In the regime of small externally applied voltage $V$, we calculate the linear conductance $G$ using the Landauer formula $G = G_{Q} \sum_{k_y} T(\eF, k_y)$, where $G_{Q} = \frac {e^2}{2\pi \hbar}$ is the conductance quantum \cite{NazarovBlanter2009}.
In the absence of the domain wall, the conductance for transport along the $x$-direction is $G_0 = G_Q \ \frac{W \eF}{\pi \hbar v}$ and $W$ is the transverse dimension of the ballistic contact.

\newcommand {\dGM} {\delta G_{\rm M}}

\begin{figure}[t]
\centering\includegraphics[width=8.4cm]{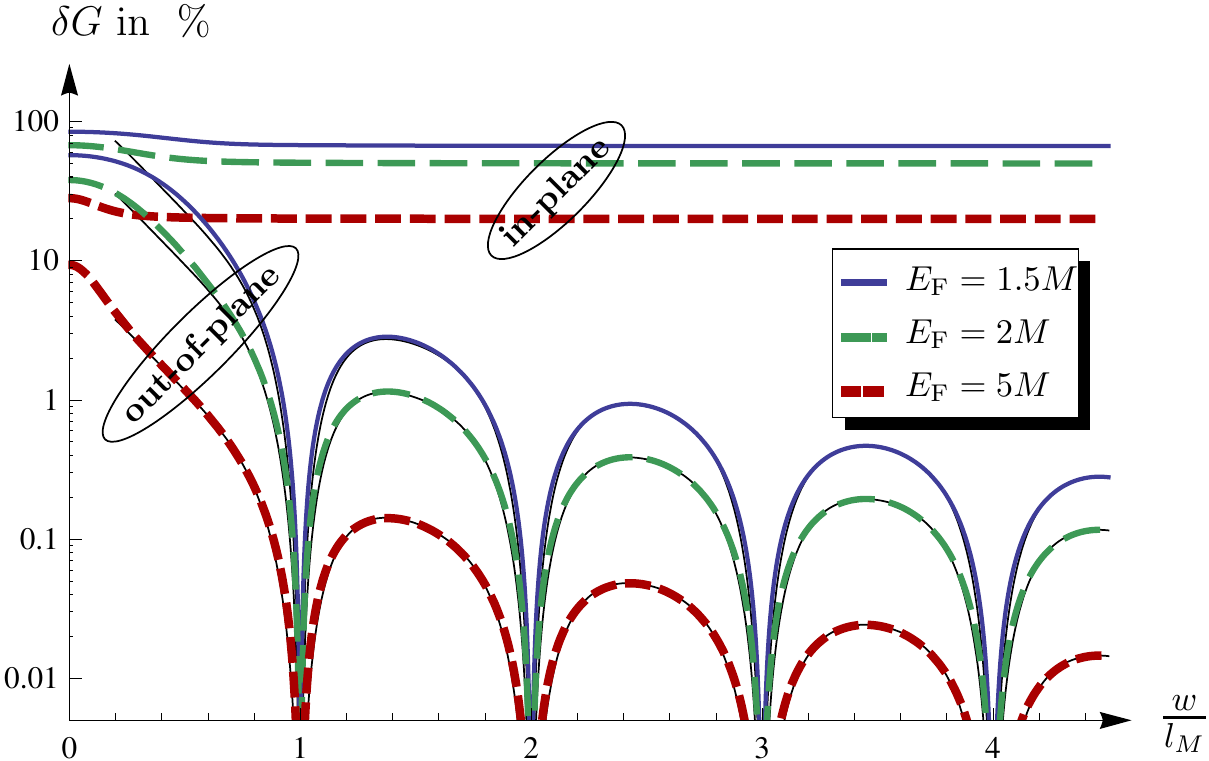}
\caption{\label{fig:DomainWallConductanceTotalEFPlot} Relative decrease in the conductance $\delta G$ due to the presence of the domain wall for in- and out-of-plane DWs. In the case of an in-plane DW, we see that the values are rather large even for wider walls and decrease monotonically towards the asymptotic value $\dGM$ as the wall width increases. To the contrary, for out-of-plane DWs it decreases significantly when the wall width becomes of the order of the magnetic length $l_M = \hbar v/M$. The approximate formula (\ref{eq:DWOOPConductanceWideWall}) is drawn as black thin lines and shows very good agreement with the exact integration. }
\end{figure}

To quantify the change of the conductance due to the presence of the domain wall, we define the domain wall resistance as
\begin{align}
\label{eq:DefinitionDWResistance}
\delta G = -\frac{G_{\rm DW} - G_0} {G_0} = \dGM + \delta G_{\rm DW} \,.
\end{align}
We split this into two contributions: $\delta G_{\rm DW}$ depends on the specific domain wall profile while $\dGM = \frac M {E_{\rm F}} > 0$ is the fraction of totally reflecting channels (red shaded area in Fig. \ref{fig:SpectrumDW}a) to the total number of transport channels (yellow + red areas). Essentially, $\dGM$ encodes the spectrum mismatch between both sides of the wall, as far as it concerns transport at the Fermi level.
The remaining contribution from the DW is then related to the reflection from the wall profile and is explicitly given by $ \delta G_{\rm DW} =  \frac 1 {2 E_{\rm F}}  \int_{-(\eF - M)}^{\eF - M} \upd k_y \  R_{\rm IP}(\eF, k_y) $. The total change in conductance $\delta G$ is shown in Fig. \ref{fig:DomainWallConductanceTotalEFPlot}, and we find that $\dGM$ strongly dominates, except for short walls for which $w \lesssim l_M$. As the Fermi level moves deeper into the metallic regime, we also see an overall decrease of the change in conductance, since the total number of transport channels increases linearly with $E_F$, while the influence of the domain-wall remains constant (e.g. the number of totally reflecting channels is determined by $M$ alone). A related study of the ballistic domain-wall resistance in the spherical Kohn-Luttinger model reveals quantitatively similar properties \cite{Nguyen2006}, in particular, they also find that $\delta G_M$ (which they call intrinsic domain-wall resistance) dominates except for sharp domain-walls.

The limiting behaviors can be obtained analytically. For abrupt walls, i.e. $w \ll l_M$, we can approximately solve the integral and find
$\delta G_{\rm DW} \stackrel{w \ll l_M}{\rightarrow} \frac 13 \dGM $. Since $\delta G_{\rm DW}$ decreases when the wall width increases, we can conclude $\delta G_{\rm DW} \le \frac 13 \dGM$. In the  limit of wide walls, i.e. $w \gg l_M$, using  the saddle point method, we find
\begin{align}
\label{eq:DWIPConductanceWideWall}
\delta G_{\rm DW} &\stackrel{w \gg l_M}{\rightarrow}  \frac {\hbar^2 v^2}{(2\pi \eF)^2} \ \frac 1{w^2} \,.
\end{align}
The ballistic domain wall resistance $\delta G_{\rm DW}$ decays with the inverse square of the domain wall width, but the contribution $\dGM = \frac M {E_{\rm F}}$ is also present in this limit and is the dominating one. Experimentally, one could subtract $\dGM$ (for example determined for an adiabatic DW) in order to obtain the contribution from the domain wall profile.

As the Fermi level comes close to the band edge $\eF \approx M$, we have $\delta G \rightarrow 1$, i.e. the domain wall blocks all transport channels. However, as discussed above, it is then no longer valid to neglect the contribution from the evanescent modes.

\section{The out-of-Plane Configuration}
The OOP wall has the magnetization profile $\vect m(\vect r) = \left(0, \sin\vartheta(x), \cos\vartheta(x)\right)$ which, in the language of the 2D-Dirac equation, describes a mass domain wall connecting two quantum anomalous Hall states of opposite chirality. We perform a spin rotation around the $y$-axis by $\frac\pi2$ utilizing the unitary spin rotation matrix $\calU = e^{-i\frac\pi4 \sigma_y}$ to obtain the representation $\calU \, H_{\rm OOP} \, \calU^\dagger = H_{\rm IP}(k_y = 0)  - \hbar v k_y \sigma_z $, which allows us to reuse the previous results.
As inferred from the symmetry $\left\{ H_{\rm IP},  \sigma_z \right\} = 0$, we see that $k_y \sigma_z$ only couples pairs of positive and negative energy.
Thus, we only have to diagonalize $2\times2$ sub-blocks and straightforwardly obtain the full energy dispersion $ E_{\vect k} = \pm \sqrt{M^2 + \hbar^2 v^2 (k_x^2 + k_y^2)}$ with corresponding eigenstates not presented here.
The same applies to the bound states except for the zero energy state for $k_{y}=0$, which is invariant under the operation of $\sigma_z$, and thus directly yields the linearly dispersing chiral state plotted as black straight line in Fig. \ref{fig:SpectrumDW}b.

In the asymptotic expansion of $\ket{\psi_{k_x,k_y}}$ far away from the domain wall, we find that finite $k_y$ only modifies the spinor structure and therefore the transmission coefficients remain independent of $k_y$. Thus, the transmission probability can be directly obtained from (\ref{eq:TransmissionProbabilityDWIP}) by setting $k_\Lid = k_\Rid = k_x$, viz.,
\begin{align}
\label{eq:DWOOPTransmissionProbability}
T_{\rm OOP}(k_x) &= \frac{\sinh^2(\pi k_x w)}{\sin^2(\pi \Delta) + \sinh^2(\pi k_x w)} \,.
\end{align}
We observe that $T_{\rm OOP}$ features oscillations in $w$ with period $l_{\rm M}$ and, in particular, for DWs with $\Delta \in \mathbb{N}$, i.e. the reflection is completely suppressed for any $k_x$ and $k_y$.

The ballistic domain wall resistance reads
$ \delta G_{\rm OOP} =\frac1{2 k_{\rm F}} \int_{-k_{\rm F}}^{k_{\rm F}} \upd k_y \ R_{\rm OOP} (\sqrt{k_{\rm F}^2 - k_y^2} ) $
with the Fermi wave-vector $\hbar v k_{\rm F} = \sqrt{E_{\rm F}^2 - M^2}$. Since the spectrum is identical on both sides of the wall, $\dGM=0$ here. This result is plotted in Fig. \ref{fig:DomainWallConductanceTotalEFPlot} and we recognize the oscillations in $\Delta$ originating from $T_{\rm OOP}$. For the special points where the domain wall width is an integer multiple of the magnetic length, we find the domain wall to be completely transparent for the Dirac Fermions and $\delta G$ drops to zero.

When the wall is much shorter than any other relevant transport length scales of the system, i.e. $\Delta, k_{\rm F} w \ll 1$, we can expand the sin in Eq.~(\ref{eq:DWOOPTransmissionProbability}) and perform the integration analytically,
\begin{align}
\label{eq:DeltaGDWshortWallAnalytic}
\delta G_{\rm OOP} \stackrel{w \ll l_{\rm M}}{\approx} \frac{M^2}{2 \hbar v k_{\rm F} E_{\rm F}} \, \log\left(\frac{E_{\rm F} + \hbar v k_{\rm F}}{E_{\rm F} - \hbar v k_{\rm F}}\right) \ .
\end{align}
For $\eF \gg M$, we find the asymptotic behavior $\delta G_{\rm OOP} \rightarrow (M/\eF)^{2}\log(2\eF/M)$, while in the opposite case when the Fermi level comes close to the band edge $\eF \approx M$, we have $\delta G_{\rm OOP} \rightarrow 1$, as for the IP-configuration.

In the limit of wide walls $\Delta, k_F w \gg 1$, we obtain the approximate result
\begin{align}
\label{eq:DWOOPConductanceWideWall}
\delta G_{\rm OOP} &\stackrel{\Delta \gg 1}{\rightarrow} \frac{\log 2} {(\pi k_{F}w)^2} \calF(\Delta)\ ,
\end{align}
with the modulation function $\calF(\Delta) = \frac {2 \sin^2(\pi\Delta)} {1+ \sin^2(\pi\Delta) }$ which inherits the periodicity from $T_{\rm OOP}$, in particular, $\calF(\Delta)$ vanishes for integer $\Delta$. We see that the envelope decreases with $1/w^2$, similar to the result for the IP DW, Eq.~(\ref{eq:DWIPConductanceWideWall}).
We find that the approximate formula fits well already for $\Delta \gtrsim 0.5$ and show in Fig. \ref{fig:DomainWallConductanceTotalEFPlot} the comparison between the full integration and the approximation (\ref{eq:DWOOPConductanceWideWall}).

In a realistic system, there is always impurity scattering which mixes channels with different $k$ and therefore reduces $\dGM$. However, as long as the wall width is much smaller than the mean free path, i.e. $w \ll l_{\rm mfp}$, our ballistic treatment is approximately correct. Impurities induce a finite scattering between the channels, however, due to the chiral nature of the electron dispersion, back-scattering is reduced. Thus, in the OOP case and when the DW conductance $\delta G$ is vanishing for reflectionless DW potentials, we expect $\delta G$ to remain significantly reduced in comparison to reflecting DWs for $\Delta \not\in \mathbbm{N}$.

\section{Supersymmetry}
The special characteristics of the domain wall resistance in the OOP geometry - the periodicity in $\Delta$ and the perfect transmission - can be understood as a supersymmetry encoded in the Hamiltonian $H_{\rm OOP}$.
Using the language of supersymmetry \cite{SuperSymmBook2001,SchwablQM2008}, we introduce the generalized creation $a^\dagger_\Delta = -\partial_\xi + \Delta \tanh(\xi)$ and annihilation operators $a_\Delta = \partial_\xi + \Delta \tanh(\xi)$ with dimensionless $\xi = x/w$. Introducing spin raising and lowering matrices $\sigma_{\pm} = (\sigma_x \pm i \sigma_y)/2$, the supercharges \cite{SuperSymmBook2001} are $Q_\Delta = a_\Delta \sigma_+$ and $Q_\Delta^\dagger = a_\Delta^\dagger \sigma_-$, so that the OOP Hamiltonian reads $\calH = \calU H_{\rm OOP} \calU^\dagger = -Q_\Delta - Q_\Delta^\dagger + k_y \sigma_z$.

We observe that the operator defined by $\bar\calH = \{Q_\Delta, Q^\dagger_\Delta \} = \frac12\{ a_\Delta, a_\Delta^\dagger \} + \frac12[ a_\Delta, a_\Delta^\dagger ] \sigma_z = \calH^2 - k_y^2$ is diagonal and essentially the square of our original Hamiltonian. Solving the eigenvalue equation for $\bar\calH$ is equivalent to the one for $\calH$. Note that in this representation, $\frac12[ a_\Delta, a_\Delta^\dagger ] \sigma_z$ removes the zero-point energy, thus allowing for the zero energy state, see Fig.~\ref{fig:SpectrumDW}. Obviously, $[\bar\calH, Q_\Delta ] = 0$, which expresses the supersymmetry between the two components $\varphi_{\UPDOWN}$ of the spinor. Explicitly, $\bar\calH_\UP = a_\Delta a_\Delta^\dagger$ and $\bar\calH_\DOWN = a_\Delta^\dagger a_\Delta$ are supersymmetric partner Hamiltonians, which are iso-spectral, except that $\bar\calH_\UP$ has one additional bound state. Furthermore, the reflection and transmission coefficients defined by $\bar\calH_\UPDOWN$ differ only by a phase \cite{SuperSymmBook2001}.

For the $\tanh$-DW profile, a second symmetry exists, $[\bar\calH - \Delta^2, \sigma_\pm e^{\mp \partial_\Delta}] = 0$. This means that  $\bar\calH_{\UPDOWN}$ are part of a hierarchy of form-invariant supersymmetric partner Hamiltonians, each differing from its neighbor by $\Delta \rightarrow \Delta + 1$. All Hamiltonians in this hierarchy have the same transmission-/reflection probabilities which readily explains the oscillations in $\delta G_{\rm OOP}$ as $w/l_M$ varies. Due to the scaling in $\xi = x/w$, there is an additional smooth dependence on $w$ which yields the factor $1/w^2$ in $\delta G_{\rm OOP}$, Eq. (\ref{eq:DWOOPConductanceWideWall}). Furthermore, if $\Delta \in \mathbbm{N}$, the constant potential is part of the hierarchy and thus all Hamiltonians in the hierarchy are reflectionless as well. Note, however, that the constant potential is not realized in our system, since the scaling becomes singular for $w\to0$. Finally, the number of bound states differs by $1$ between two neighbors in the hierarchy, which explains that the number of bound states is given by $\lfloor\Delta\rfloor$. We point out that similar reasoning can be used for the IP wall configuration, there however, the supersymmetric hierarchy is constructed in $(\Delta, k_y)$-space and thus, the $k_y$-integration performed in $\delta G$ averages out the characteristic signature of the hierarchy.

We remark that $\bar\calH_\UPDOWN$ describes essentially a free particle in the P\"oschl-Teller potential, which is known to be reflectionless for certain parameters. In the reflectionless case however, a transmitted wave still acquires a phase which has consequences on a wave packet passing such a potential: it narrows and is ahead in time as compared to a freely moving wave-packet \cite{Kiriushcheva1998}.
For optical systems, these potentials have been realized recently using arrays of evanescently coupled waveguides \cite{Szameit2011}.

Finally we comment briefly on experimental issues. A possible realization is to deposit an isolating ferromagnetic film on top of the topological insulator or alternatively, a metallic ferromagnet with a thin insulating barrier. While there is a broad range of tunable magnetic properties in metallic materials \cite{Fassbender2004} which allows one to create domain walls of desired size and configuration, for isolating ferromagnets the situation is more difficult, but recent experiments show that challenges like perpendicular magnetization are feasible in such films \cite{Xia2010}. The observation of the predicted oscillations could be realized by placing an additional top-gate on the ferromagnetic insulator which can be used to tune the exchange field $M$. The oscillations would still be visible if the ratio $E_F/M$ changes slowly enough.

\section{Conclusions}
We have analytically calculated the ballistic DW conductance for in-plane and out-of-plane magnetic domain walls induced into the surface states of a topological insulator.
For the in-plane DW, the DW conductance is dominated by the spectrum mismatch imposed by the opposite magnetization directions within the  domains.
For the out-of-plane DW, we unexpectedly find oscillations in the wall-width dependence with period $l_M$ (magnetic length). In particular, integer $w/l_M$ constitute a family of reflectionless potentials. We can understand these features using the idea of supersymmetry and find them to be a result of the dispersion of the topological surface states together with the specific tanh DW-profile. Detecting the oscillatory DW resistance could be a unique signature of the chiral Dirac surface states.

We would like to thank Christoph Bruder, Mathias Kl\"aui and Cord A. M\"uller for helpful discussions. This work was financially supported by the DFG though SFB 767 and SP1285.

\appendix
\section{Hypergeometric Function}

The hypergeometric function $_2F_1(\alpha,\beta;\gamma;z)$ is defined as \cite{Gradshteyn2007}
\begin{alignat}{2}
\label{eq:HypergeometricDefinition}
_2F_1(\alpha,\beta;\gamma;z) &= \sum_{n=0}^\infty \frac{(\alpha)_n (\beta)_n}{(\gamma)_n} \, \frac {z^n} {n!} \ ,
\end{alignat}
where $(\alpha)_n$ is the Pochhammer symbol
\begin{align}
(\alpha)_n = \frac{\Gamma(\alpha+n)}{\Gamma(\alpha)} \ .
\end{align}
The Hypergeometric function has a branch cut on the real axis for $\Re z >=1$, and the series expansion for $z \rightarrow 1$ and $\gamma - \alpha - \beta \not\in Z$, $\Re(\gamma-\alpha-\beta) \ge 0$ reads ($\xi > 0$)
\begin{multline}
\label{eq:HypergeometricExpansion}
_2F_1(\alpha,\beta;\gamma;1-\xi) =  \frac{\Gamma(\gamma) \Gamma(\gamma - \alpha - \beta)}{\Gamma(\gamma - \alpha) \Gamma(\gamma - \beta)} \left( 1 + O(\xi) \right) \\
 + \frac{\Gamma(\gamma) \Gamma(\alpha + \beta - \gamma)}{\Gamma(\alpha) \Gamma(\beta)} \xi^{\gamma - \alpha - \beta} \left( 1 + O(\xi) \right) \ .
\end{multline}
In these expressions, $\Gamma(z)$ is the Gamma function \cite{Gradshteyn2007} with some properties useful for the derivation of result (\ref{eq:TransmissionProbabilityDWIP}) given,
\begin{align}
\Gamma(z+1) &= z \Gamma(z) \\
\Gamma(z) \Gamma(-z) &= -\frac{\pi}{z\sin(\pi z)} \\
\Gamma(2z) &= \frac{2^{2z-1}}{\sqrt\pi} \Gamma(z) \Gamma(z+\frac12) \ .
\end{align}

\section{Asymptotic Expansion and Bound States}
To set the stage, we specify the spin eigenstates of the bulk system on the left and right sides of the domain wall,
\begin{align}
\label{eq:ChiLRDefinition}
\ket{\chi_\LRidx} = \frac 1{\sqrt2} \begin{pmatrix} e^{-\frac i2 (\gamma_\LRidx - \frac\pi2)} \\ e^{+\frac i2 (\gamma_\LRidx - \frac\pi2)} \end{pmatrix} \ .
\end{align}
In this state, the current in the transport direction is then simply given by
\begin{align}
\label{eq:CurrentValueChiLR}
\bra{\chi_\LRidx} j_x \ket{\chi_\LRidx} = v \Re e^{i \gamma_\LRidx} = v  \frac {k_\LRid} {\epsilon} \ .
\end{align}

By making a series expansion of the hypergeometric function around $z=1$ ($x \rightarrow -\infty$) and $z=0$ ($x \rightarrow +\infty$) (see series expansions (\ref{eq:HypergeometricDefinition}) and (\ref{eq:HypergeometricExpansion})), the asymptotics of $\varphi_\UP(z)$ given in Eq. (\ref{eq:HypergeometricSolutionPhiUp}) and $\varphi_\DOWN(z)=-i \calT_M \calT_{k_y}\varphi_\UP(z)$ can be compactly written as
\begin{widetext}
\begin{align}
\label{eq:WavefunctionAsymptoticsincomingLeft}
\ket{\psi_{k_\Lid,k_y}^{\rm (IP)}} = \begin{cases} A(k_\Lid, k_\Rid) \dfrac{e^{i k_\Lid x}}{\sqrt {v_\Lid}} \ket{\chi_\Lidx} - A(-k_\Lid, k_\Rid) \dfrac{e^{-i k_\Lid x}}{\sqrt {v_\Lid}} \ket{\chi_\Lidx^*} & x \rightarrow -\infty \\
\dfrac{e^{i k_\Rid x}}{\sqrt {v_\Rid}} \ket{\chi_\Ridx} & x \rightarrow +\infty \ .
\end{cases}
\end{align}
Note that in order to obtain properly normalized reflection and transmission amplitudes, we need to include the group-velocity in transport direction in the prefactor, since $v_\LRid$ is different on left and right sides of the wall.
As mentioned in the main text, the second independent solution can be most easily found using $-i \calT_{k_y} \calT_x \sigma_z \ket{\psi_{k_\Lid,k_y}^{\rm (IP)}}$, so that
\begin{align}
\label{eq:WavefunctionAsymptoticsincomingRight}
\ket{\psi_{-k_\Rid,k_y}^{\rm (IP)}} = \begin{cases} \dfrac{e^{-i k_\Lid x}}{\sqrt {v_\Lid}} \ket{\chi_\Lidx^*} & x \rightarrow -\infty \\
 A(k_\Lid, k_\Rid) \dfrac{e^{- i k_\Rid x}}{\sqrt {v_\Rid}} \ket{\chi_\Ridx^*} + A(k_\Lid, -k_\Rid) \dfrac{e^{i k_\Rid x}}{\sqrt {v_\Rid}} \ket{\chi_\Ridx} & x \rightarrow +\infty \ ,
\end{cases}
\end{align}
where we employed $\calT_{k_y} k_\LRidx = k_\RLidx$, $\calT_{k_y} e^{i\gamma_\LRidx} = e^{-i\gamma_\RLidx}$ and $\calT_{k_y} e^{-i\frac\pi2} \sigma_z \ket{\chi_\LRidx} = \ket{\chi_\RLidx^*}$. Furthermore, we defined the coefficient
\begin{align}
\label{eq:AkLkRDefinition}
A(k_\Lid, k_\Rid) = \frac {\sqrt{\frac14 \left(k_\Lid + k_\Rid\right)^2 w^2 + \Delta^2}} {w\sqrt{\left|k_\Lid k_\Rid\right|}}  \frac{\Gamma(1-ik_\Lid w)\Gamma(1-ik_\Rid w)}{\Gamma(1-\Delta-\frac i2 (k_\Lid + k_\Rid) w) \Gamma(1+\Delta-\frac i2 (k_\Lid + k_\Rid) w)}
\end{align}
with the properties $A(k_\Lid, k_\Rid)^* = A(-k_\Lid, -k_\Rid)$, $\calT_{k_y} A(k_\Lid, k_\Rid) = A(k_\Rid, k_\Lid) = A(k_\Lid, k_\Rid)$ and $\calT_M A(k_\Lid, k_\Rid) = A(k_\Lid, k_\Rid)$.
\end{widetext}

From (\ref{eq:WavefunctionAsymptoticsincomingLeft}), one readily retrieves the reflection and transmission amplitudes for $\ket{\psi_{k_\Lid,k_y}^{\rm (IP)}}$,
\begin{align}
\label{eq:ReflectionCoefficientDWXY}
r &= \frac {-A(-k_\Lid, k_\Rid)} {A(k_\Lid, k_\Rid)} \ , \\
\label{eq:TransmissionCoefficientDWXY}
t &= \frac {1} {A(k_\Lid, k_\Rid)} \ ,
\end{align}
from which one can straightforwardly find result (\ref{eq:TransmissionProbabilityDWIP}).

We are now looking for bound states, i.e. solutions with imaginary wave-vector $-i k_\RLid > 0$, so that the wave function becomes exponentially localized at the domain wall. We can easily investigate this with the help of the asymptotic expansion (\ref{eq:WavefunctionAsymptoticsincomingLeft}), and since $\frac{1}{\Gamma(-n)} = 0$ for $n \ge 0$, we find from definition (\ref{eq:AkLkRDefinition}) that $A(k_\Lid, k_\Rid) = 0$, provided that
\begin{align}
-\frac i2(k_\Lid + k_\Rid) w = \Delta -n \ , \quad n = 0,1,2 \dots \lfloor\Delta \rfloor \ .
\end{align}
This condition immediately transforms into the dispersion relation $E_{n,k_y}$ given in the main text. Finally, we remark that for integer $n$, the hypergeometric series terminates and one has a polynomial in $z$, {\it viz.} $\tanh(x/w)$.

%


\begin{thebibliography}{29}%
\makeatletter
\providecommand \@ifxundefined [1]{%
 \@ifx{#1\undefined}
}%
\providecommand \@ifnum [1]{%
 \ifnum #1\expandafter \@firstoftwo
 \else \expandafter \@secondoftwo
 \fi
}%
\providecommand \@ifx [1]{%
 \ifx #1\expandafter \@firstoftwo
 \else \expandafter \@secondoftwo
 \fi
}%
\providecommand \natexlab [1]{#1}%
\providecommand \enquote  [1]{``#1''}%
\providecommand \bibnamefont  [1]{#1}%
\providecommand \bibfnamefont [1]{#1}%
\providecommand \citenamefont [1]{#1}%
\providecommand \href@noop [0]{\@secondoftwo}%
\providecommand \href [0]{\begingroup \@sanitize@url \@href}%
\providecommand \@href[1]{\@@startlink{#1}\@@href}%
\providecommand \@@href[1]{\endgroup#1\@@endlink}%
\providecommand \@sanitize@url [0]{\catcode `\\12\catcode `\$12\catcode
  `\&12\catcode `\#12\catcode `\^12\catcode `\_12\catcode `\%12\relax}%
\providecommand \@@startlink[1]{}%
\providecommand \@@endlink[0]{}%
\providecommand \url  [0]{\begingroup\@sanitize@url \@url }%
\providecommand \@url [1]{\endgroup\@href {#1}{\urlprefix }}%
\providecommand \urlprefix  [0]{URL }%
\providecommand \Eprint [0]{\href }%
\providecommand \doibase [0]{http://dx.doi.org/}%
\providecommand \selectlanguage [0]{\@gobble}%
\providecommand \bibinfo  [0]{\@secondoftwo}%
\providecommand \bibfield  [0]{\@secondoftwo}%
\providecommand \translation [1]{[#1]}%
\providecommand \BibitemOpen [0]{}%
\providecommand \bibitemStop [0]{}%
\providecommand \bibitemNoStop [0]{.\EOS\space}%
\providecommand \EOS [0]{\spacefactor3000\relax}%
\providecommand \BibitemShut  [1]{\csname bibitem#1\endcsname}%
\let\auto@bib@innerbib\@empty
\bibitem [{\citenamefont {Hsieh}\ \emph {et~al.}(2008)\citenamefont {Hsieh},
  \citenamefont {Qian}, \citenamefont {Wray}, \citenamefont {Xia},
  \citenamefont {Hor}, \citenamefont {Cava},\ and\ \citenamefont
  {Hasan}}]{Hsieh2008}%
  \BibitemOpen
  \bibfield  {author} {\bibinfo {author} {\bibfnamefont {D.}~\bibnamefont
  {Hsieh}}, \bibinfo {author} {\bibfnamefont {D.}~\bibnamefont {Qian}},
  \bibinfo {author} {\bibfnamefont {L.}~\bibnamefont {Wray}}, \bibinfo {author}
  {\bibfnamefont {Y.}~\bibnamefont {Xia}}, \bibinfo {author} {\bibfnamefont
  {Y.~S.}\ \bibnamefont {Hor}}, \bibinfo {author} {\bibfnamefont {R.~J.}\
  \bibnamefont {Cava}}, \ and\ \bibinfo {author} {\bibfnamefont {M.~Z.}\
  \bibnamefont {Hasan}},\ }\href {http://dx.doi.org/10.1038/nature06843}
  {\bibfield  {journal} {\bibinfo  {journal} {Nature}\ }\textbf {\bibinfo
  {volume} {452}},\ \bibinfo {pages} {970} (\bibinfo {year}
  {2008})}\BibitemShut {NoStop}%
\bibitem [{\citenamefont {Zhang}\ \emph {et~al.}(2009)\citenamefont {Zhang},
  \citenamefont {Liu}, \citenamefont {Qi}, \citenamefont {Dai}, \citenamefont
  {Fang},\ and\ \citenamefont {Zhang}}]{Zhang2009}%
  \BibitemOpen
  \bibfield  {author} {\bibinfo {author} {\bibfnamefont {H.}~\bibnamefont
  {Zhang}}, \bibinfo {author} {\bibfnamefont {C.-X.}\ \bibnamefont {Liu}},
  \bibinfo {author} {\bibfnamefont {X.-L.}\ \bibnamefont {Qi}}, \bibinfo
  {author} {\bibfnamefont {X.}~\bibnamefont {Dai}}, \bibinfo {author}
  {\bibfnamefont {Z.}~\bibnamefont {Fang}}, \ and\ \bibinfo {author}
  {\bibfnamefont {S.-C.}\ \bibnamefont {Zhang}},\ }\href
  {http://dx.doi.org/10.1038/nphys1270} {\bibfield  {journal} {\bibinfo
  {journal} {Nat Phys}\ }\textbf {\bibinfo {volume} {5}},\ \bibinfo {pages}
  {438} (\bibinfo {year} {2009})}\BibitemShut {NoStop}%
\bibitem [{\citenamefont {Fu}\ and\ \citenamefont {Kane}(2009)}]{FuKane2009}%
  \BibitemOpen
  \bibfield  {author} {\bibinfo {author} {\bibfnamefont {L.}~\bibnamefont
  {Fu}}\ and\ \bibinfo {author} {\bibfnamefont {C.~L.}\ \bibnamefont {Kane}},\
  }\href {\doibase 10.1103/PhysRevLett.102.216403} {\bibfield  {journal}
  {\bibinfo  {journal} {Phys. Rev. Lett.}\ }\textbf {\bibinfo {volume} {102}},\
  \bibinfo {pages} {216403} (\bibinfo {year} {2009})}\BibitemShut {NoStop}%
\bibitem [{\citenamefont {Akhmerov}\ \emph {et~al.}(2009)\citenamefont
  {Akhmerov}, \citenamefont {Nilsson},\ and\ \citenamefont
  {Beenakker}}]{Akhmerov2009}%
  \BibitemOpen
  \bibfield  {author} {\bibinfo {author} {\bibfnamefont {A.~R.}\ \bibnamefont
  {Akhmerov}}, \bibinfo {author} {\bibfnamefont {J.}~\bibnamefont {Nilsson}}, \
  and\ \bibinfo {author} {\bibfnamefont {C.~W.~J.}\ \bibnamefont {Beenakker}},\
  }\href {\doibase 10.1103/PhysRevLett.102.216404} {\bibfield  {journal}
  {\bibinfo  {journal} {Phys. Rev. Lett.}\ }\textbf {\bibinfo {volume} {102}},\
  \bibinfo {pages} {216404} (\bibinfo {year} {2009})}\BibitemShut {NoStop}%
\bibitem [{\citenamefont {Tanaka}\ \emph {et~al.}(2009)\citenamefont {Tanaka},
  \citenamefont {Yokoyama},\ and\ \citenamefont {Nagaosa}}]{Tanaka2009}%
  \BibitemOpen
  \bibfield  {author} {\bibinfo {author} {\bibfnamefont {Y.}~\bibnamefont
  {Tanaka}}, \bibinfo {author} {\bibfnamefont {T.}~\bibnamefont {Yokoyama}}, \
  and\ \bibinfo {author} {\bibfnamefont {N.}~\bibnamefont {Nagaosa}},\ }\href
  {\doibase 10.1103/PhysRevLett.103.107002} {\bibfield  {journal} {\bibinfo
  {journal} {Phys. Rev. Lett.}\ }\textbf {\bibinfo {volume} {103}},\ \bibinfo
  {pages} {107002} (\bibinfo {year} {2009})}\BibitemShut {NoStop}%
\bibitem [{\citenamefont {Linder}\ \emph {et~al.}(2010)\citenamefont {Linder},
  \citenamefont {Tanaka}, \citenamefont {Yokoyama}, \citenamefont {Sudb\o{}},\
  and\ \citenamefont {Nagaosa}}]{Linder2010a}%
  \BibitemOpen
  \bibfield  {author} {\bibinfo {author} {\bibfnamefont {J.}~\bibnamefont
  {Linder}}, \bibinfo {author} {\bibfnamefont {Y.}~\bibnamefont {Tanaka}},
  \bibinfo {author} {\bibfnamefont {T.}~\bibnamefont {Yokoyama}}, \bibinfo
  {author} {\bibfnamefont {A.}~\bibnamefont {Sudb\o{}}}, \ and\ \bibinfo
  {author} {\bibfnamefont {N.}~\bibnamefont {Nagaosa}},\ }\href {\doibase
  10.1103/PhysRevLett.104.067001} {\bibfield  {journal} {\bibinfo  {journal}
  {Phys. Rev. Lett.}\ }\textbf {\bibinfo {volume} {104}},\ \bibinfo {pages}
  {067001} (\bibinfo {year} {2010})}\BibitemShut {NoStop}%
\bibitem [{\citenamefont {Qi}\ \emph {et~al.}(2008{\natexlab{a}})\citenamefont
  {Qi}, \citenamefont {Hughes},\ and\ \citenamefont
  {Zhang}}]{QiXiaoTopologicalFieldTheory2008}%
  \BibitemOpen
  \bibfield  {author} {\bibinfo {author} {\bibfnamefont {X.-L.}\ \bibnamefont
  {Qi}}, \bibinfo {author} {\bibfnamefont {T.~L.}\ \bibnamefont {Hughes}}, \
  and\ \bibinfo {author} {\bibfnamefont {S.-C.}\ \bibnamefont {Zhang}},\ }\href
  {\doibase 10.1103/PhysRevB.78.195424} {\bibfield  {journal} {\bibinfo
  {journal} {Phys. Rev. B}\ }\textbf {\bibinfo {volume} {78}},\ \bibinfo
  {pages} {195424} (\bibinfo {year} {2008}{\natexlab{a}})}\BibitemShut
  {NoStop}%
\bibitem [{\citenamefont {Qi}\ \emph {et~al.}(2008{\natexlab{b}})\citenamefont
  {Qi}, \citenamefont {Hughes},\ and\ \citenamefont {Zhang}}]{Qi2008NAT}%
  \BibitemOpen
  \bibfield  {author} {\bibinfo {author} {\bibfnamefont {X.-L.}\ \bibnamefont
  {Qi}}, \bibinfo {author} {\bibfnamefont {T.~L.}\ \bibnamefont {Hughes}}, \
  and\ \bibinfo {author} {\bibfnamefont {S.-C.}\ \bibnamefont {Zhang}},\ }\href
  {http://dx.doi.org/10.1038/nphys913} {\bibfield  {journal} {\bibinfo
  {journal} {Nat Phys}\ }\textbf {\bibinfo {volume} {4}},\ \bibinfo {pages}
  {273} (\bibinfo {year} {2008}{\natexlab{b}})}\BibitemShut {NoStop}%
\bibitem [{\citenamefont {Qi}\ and\ \citenamefont
  {Zhang}(2011)}]{QiZhangRMP2011}%
  \BibitemOpen
  \bibfield  {author} {\bibinfo {author} {\bibfnamefont {X.-L.}\ \bibnamefont
  {Qi}}\ and\ \bibinfo {author} {\bibfnamefont {S.-C.}\ \bibnamefont {Zhang}},\
  }\href {\doibase 10.1103/RevModPhys.83.1057} {\bibfield  {journal} {\bibinfo
  {journal} {Rev. Mod. Phys.}\ }\textbf {\bibinfo {volume} {83}},\ \bibinfo
  {pages} {1057} (\bibinfo {year} {2011})}\BibitemShut {NoStop}%
\bibitem [{\citenamefont {Mondal}\ \emph {et~al.}(2010)\citenamefont {Mondal},
  \citenamefont {Sen}, \citenamefont {Sengupta},\ and\ \citenamefont
  {Shankar}}]{Mondal2010}%
  \BibitemOpen
  \bibfield  {author} {\bibinfo {author} {\bibfnamefont {S.}~\bibnamefont
  {Mondal}}, \bibinfo {author} {\bibfnamefont {D.}~\bibnamefont {Sen}},
  \bibinfo {author} {\bibfnamefont {K.}~\bibnamefont {Sengupta}}, \ and\
  \bibinfo {author} {\bibfnamefont {R.}~\bibnamefont {Shankar}},\ }\href
  {\doibase 10.1103/PhysRevLett.104.046403} {\bibfield  {journal} {\bibinfo
  {journal} {Phys. Rev. Lett.}\ }\textbf {\bibinfo {volume} {104}},\ \bibinfo
  {pages} {046403} (\bibinfo {year} {2010})}\BibitemShut {NoStop}%
\bibitem [{\citenamefont {Garate}\ and\ \citenamefont
  {Franz}(2010)}]{Garate2010}%
  \BibitemOpen
  \bibfield  {author} {\bibinfo {author} {\bibfnamefont {I.}~\bibnamefont
  {Garate}}\ and\ \bibinfo {author} {\bibfnamefont {M.}~\bibnamefont {Franz}},\
  }\href {\doibase 10.1103/PhysRevLett.104.146802} {\bibfield  {journal}
  {\bibinfo  {journal} {Phys. Rev. Lett.}\ }\textbf {\bibinfo {volume} {104}},\
  \bibinfo {pages} {146802} (\bibinfo {year} {2010})}\BibitemShut {NoStop}%
\bibitem [{\citenamefont {Yokoyama}(2011)}]{Yokoyama2011}%
  \BibitemOpen
  \bibfield  {author} {\bibinfo {author} {\bibfnamefont {T.}~\bibnamefont
  {Yokoyama}},\ }\href {\doibase 10.1103/PhysRevB.84.113407} {\bibfield
  {journal} {\bibinfo  {journal} {Phys. Rev. B}\ }\textbf {\bibinfo {volume}
  {84}},\ \bibinfo {pages} {113407} (\bibinfo {year} {2011})}\BibitemShut
  {NoStop}%
\bibitem [{\citenamefont {Tserkovnyak}\ and\ \citenamefont
  {Loss}(2011)}]{Tserkovnyak2012}%
  \BibitemOpen
  \bibfield  {author} {\bibinfo {author} {\bibfnamefont {Y.}~\bibnamefont
  {Tserkovnyak}}\ and\ \bibinfo {author} {\bibfnamefont {D.}~\bibnamefont
  {Loss}},\ }\href@noop {} {\bibfield  {journal} {\bibinfo  {journal} {ArXiv
  e-prints}\ } (\bibinfo {year} {2011})},\ \Eprint
  {http://arxiv.org/abs/1112.5884} {arXiv:1112.5884 [cond-mat.mes-hall]}
  \BibitemShut {NoStop}%
\bibitem [{\citenamefont {Li}\ \emph {et~al.}(2010)\citenamefont {Li},
  \citenamefont {Wang}, \citenamefont {Qi},\ and\ \citenamefont
  {Zhang}}]{Li2010}%
  \BibitemOpen
  \bibfield  {author} {\bibinfo {author} {\bibfnamefont {R.}~\bibnamefont
  {Li}}, \bibinfo {author} {\bibfnamefont {J.}~\bibnamefont {Wang}}, \bibinfo
  {author} {\bibfnamefont {X.-L.}\ \bibnamefont {Qi}}, \ and\ \bibinfo {author}
  {\bibfnamefont {S.-C.}\ \bibnamefont {Zhang}},\ }\href
  {http://dx.doi.org/10.1038/nphys1534} {\bibfield  {journal} {\bibinfo
  {journal} {Nat Phys}\ }\textbf {\bibinfo {volume} {6}},\ \bibinfo {pages}
  {284} (\bibinfo {year} {2010})}\BibitemShut {NoStop}%
\bibitem [{\citenamefont {Yokoyama}\ and\ \citenamefont
  {Linder}(2011)}]{Yokoyama2011Graphene}%
  \BibitemOpen
  \bibfield  {author} {\bibinfo {author} {\bibfnamefont {T.}~\bibnamefont
  {Yokoyama}}\ and\ \bibinfo {author} {\bibfnamefont {J.}~\bibnamefont
  {Linder}},\ }\href {\doibase 10.1103/PhysRevB.83.081418} {\bibfield
  {journal} {\bibinfo  {journal} {Phys. Rev. B}\ }\textbf {\bibinfo {volume}
  {83}},\ \bibinfo {pages} {081418} (\bibinfo {year} {2011})}\BibitemShut
  {NoStop}%
\bibitem [{\citenamefont {Ren}\ \emph {et~al.}(2011)\citenamefont {Ren},
  \citenamefont {Taskin}, \citenamefont {Sasaki}, \citenamefont {Segawa},\ and\
  \citenamefont {Ando}}]{Ren2011}%
  \BibitemOpen
  \bibfield  {author} {\bibinfo {author} {\bibfnamefont {Z.}~\bibnamefont
  {Ren}}, \bibinfo {author} {\bibfnamefont {A.~A.}\ \bibnamefont {Taskin}},
  \bibinfo {author} {\bibfnamefont {S.}~\bibnamefont {Sasaki}}, \bibinfo
  {author} {\bibfnamefont {K.}~\bibnamefont {Segawa}}, \ and\ \bibinfo {author}
  {\bibfnamefont {Y.}~\bibnamefont {Ando}},\ }\href {\doibase
  10.1103/PhysRevB.84.075316} {\bibfield  {journal} {\bibinfo  {journal} {Phys.
  Rev. B}\ }\textbf {\bibinfo {volume} {84}},\ \bibinfo {pages} {075316}
  (\bibinfo {year} {2011})}\BibitemShut {NoStop}%
\bibitem [{\citenamefont {Hsieh}\ \emph {et~al.}(2009)\citenamefont {Hsieh},
  \citenamefont {Xia}, \citenamefont {Qian}, \citenamefont {Wray},
  \citenamefont {Dil}, \citenamefont {Meier}, \citenamefont {Osterwalder},
  \citenamefont {Patthey}, \citenamefont {Checkelsky}, \citenamefont {Ong},
  \citenamefont {Fedorov}, \citenamefont {Lin}, \citenamefont {Bansil},
  \citenamefont {Grauer}, \citenamefont {Hor}, \citenamefont {Cava},\ and\
  \citenamefont {Hasan}}]{Hsieh2009}%
  \BibitemOpen
  \bibfield  {author} {\bibinfo {author} {\bibfnamefont {D.}~\bibnamefont
  {Hsieh}}, \bibinfo {author} {\bibfnamefont {Y.}~\bibnamefont {Xia}}, \bibinfo
  {author} {\bibfnamefont {D.}~\bibnamefont {Qian}}, \bibinfo {author}
  {\bibfnamefont {L.}~\bibnamefont {Wray}}, \bibinfo {author} {\bibfnamefont
  {J.~H.}\ \bibnamefont {Dil}}, \bibinfo {author} {\bibfnamefont
  {F.}~\bibnamefont {Meier}}, \bibinfo {author} {\bibfnamefont
  {J.}~\bibnamefont {Osterwalder}}, \bibinfo {author} {\bibfnamefont
  {L.}~\bibnamefont {Patthey}}, \bibinfo {author} {\bibfnamefont {J.~G.}\
  \bibnamefont {Checkelsky}}, \bibinfo {author} {\bibfnamefont {N.~P.}\
  \bibnamefont {Ong}}, \bibinfo {author} {\bibfnamefont {A.~V.}\ \bibnamefont
  {Fedorov}}, \bibinfo {author} {\bibfnamefont {H.}~\bibnamefont {Lin}},
  \bibinfo {author} {\bibfnamefont {A.}~\bibnamefont {Bansil}}, \bibinfo
  {author} {\bibfnamefont {D.}~\bibnamefont {Grauer}}, \bibinfo {author}
  {\bibfnamefont {Y.~S.}\ \bibnamefont {Hor}}, \bibinfo {author} {\bibfnamefont
  {R.~J.}\ \bibnamefont {Cava}}, \ and\ \bibinfo {author} {\bibfnamefont
  {M.~Z.}\ \bibnamefont {Hasan}},\ }\href
  {http://dx.doi.org/10.1038/nature08234} {\bibfield  {journal} {\bibinfo
  {journal} {Nature}\ }\textbf {\bibinfo {volume} {460}},\ \bibinfo {pages}
  {1101} (\bibinfo {year} {2009})}\BibitemShut {NoStop}%
\bibitem [{\citenamefont {Tatara}\ and\ \citenamefont
  {Kohno}(2004)}]{Tatara2004}%
  \BibitemOpen
  \bibfield  {author} {\bibinfo {author} {\bibfnamefont {G.}~\bibnamefont
  {Tatara}}\ and\ \bibinfo {author} {\bibfnamefont {H.}~\bibnamefont {Kohno}},\
  }\href {\doibase 10.1103/PhysRevLett.92.086601} {\bibfield  {journal}
  {\bibinfo  {journal} {Phys. Rev. Lett.}\ }\textbf {\bibinfo {volume} {92}},\
  \bibinfo {pages} {086601} (\bibinfo {year} {2004})}\BibitemShut {NoStop}%
\bibitem [{\citenamefont {Gradshteyn}\ and\ \citenamefont
  {Ryzhik}(2007)}]{Gradshteyn2007}%
  \BibitemOpen
  \bibfield  {author} {\bibinfo {author} {\bibfnamefont {I.}~\bibnamefont
  {Gradshteyn}}\ and\ \bibinfo {author} {\bibfnamefont {I.}~\bibnamefont
  {Ryzhik}},\ }\href@noop {} {\emph {\bibinfo {title} {Table of Integrals,
  Series, and Products (Seventh Edition)}}}\ (\bibinfo  {publisher} {Academic
  Press},\ \bibinfo {year} {2007})\BibitemShut {NoStop}%
\bibitem [{\citenamefont {Katsnelson}(2006)}]{Katsnelson2006}%
  \BibitemOpen
  \bibfield  {author} {\bibinfo {author} {\bibfnamefont {M.}~\bibnamefont
  {Katsnelson}},\ }\href {\doibase 10.1140/epjb/e2006-00203-1} {\bibfield
  {journal} {\bibinfo  {journal} {Eur. Phys. J. B}\ }\textbf {\bibinfo {volume}
  {51}},\ \bibinfo {pages} {157} (\bibinfo {year} {2006})}\BibitemShut
  {NoStop}%
\bibitem [{\citenamefont {Tworzyd\l{}o}\ \emph {et~al.}(2006)\citenamefont
  {Tworzyd\l{}o}, \citenamefont {Trauzettel}, \citenamefont {Titov},
  \citenamefont {Rycerz},\ and\ \citenamefont {Beenakker}}]{Tworzydo2006}%
  \BibitemOpen
  \bibfield  {author} {\bibinfo {author} {\bibfnamefont {J.}~\bibnamefont
  {Tworzyd\l{}o}}, \bibinfo {author} {\bibfnamefont {B.}~\bibnamefont
  {Trauzettel}}, \bibinfo {author} {\bibfnamefont {M.}~\bibnamefont {Titov}},
  \bibinfo {author} {\bibfnamefont {A.}~\bibnamefont {Rycerz}}, \ and\ \bibinfo
  {author} {\bibfnamefont {C.~W.~J.}\ \bibnamefont {Beenakker}},\ }\href
  {\doibase 10.1103/PhysRevLett.96.246802} {\bibfield  {journal} {\bibinfo
  {journal} {Phys. Rev. Lett.}\ }\textbf {\bibinfo {volume} {96}},\ \bibinfo
  {pages} {246802} (\bibinfo {year} {2006})}\BibitemShut {NoStop}%
\bibitem [{\citenamefont {Nazarov}\ and\ \citenamefont
  {Blanter}(2009)}]{NazarovBlanter2009}%
  \BibitemOpen
  \bibfield  {author} {\bibinfo {author} {\bibfnamefont {Y.~V.}\ \bibnamefont
  {Nazarov}}\ and\ \bibinfo {author} {\bibfnamefont {Y.~M.}\ \bibnamefont
  {Blanter}},\ }\href@noop {} {\emph {\bibinfo {title} {Quantum Transport:
  Introduction to Nanoscience}}}\ (\bibinfo  {publisher} {Cambridge University
  Press},\ \bibinfo {year} {2009})\BibitemShut {NoStop}%
\bibitem [{\citenamefont {Nguyen}\ \emph {et~al.}(2006)\citenamefont {Nguyen},
  \citenamefont {Shchelushkin},\ and\ \citenamefont {Brataas}}]{Nguyen2006}%
  \BibitemOpen
  \bibfield  {author} {\bibinfo {author} {\bibfnamefont {A.~K.}\ \bibnamefont
  {Nguyen}}, \bibinfo {author} {\bibfnamefont {R.~V.}\ \bibnamefont
  {Shchelushkin}}, \ and\ \bibinfo {author} {\bibfnamefont {A.}~\bibnamefont
  {Brataas}},\ }\href {\doibase 10.1103/PhysRevLett.97.136603} {\bibfield
  {journal} {\bibinfo  {journal} {Phys. Rev. Lett.}\ }\textbf {\bibinfo
  {volume} {97}},\ \bibinfo {pages} {136603} (\bibinfo {year}
  {2006})}\BibitemShut {NoStop}%
\bibitem [{\citenamefont {Cooper}\ \emph {et~al.}(2001)\citenamefont {Cooper},
  \citenamefont {Khare},\ and\ \citenamefont {Sukhatme}}]{SuperSymmBook2001}%
  \BibitemOpen
  \bibfield  {author} {\bibinfo {author} {\bibfnamefont {F.}~\bibnamefont
  {Cooper}}, \bibinfo {author} {\bibfnamefont {A.}~\bibnamefont {Khare}}, \
  and\ \bibinfo {author} {\bibfnamefont {U.}~\bibnamefont {Sukhatme}},\
  }\href@noop {} {\emph {\bibinfo {title} {Supersymmetry in Quantum
  Mechanics}}}\ (\bibinfo  {publisher} {World Scientific Publishing},\ \bibinfo
  {year} {2001})\BibitemShut {NoStop}%
\bibitem [{\citenamefont {Schwabl}(2008)}]{SchwablQM2008}%
  \BibitemOpen
  \bibfield  {author} {\bibinfo {author} {\bibfnamefont {F.}~\bibnamefont
  {Schwabl}},\ }\href@noop {} {\emph {\bibinfo {title} {Advanced Quantum
  Mechanics}}},\ edited by\ \bibinfo {editor} {\bibfnamefont {R.}~\bibnamefont
  {Hilton}}\ and\ \bibinfo {editor} {\bibfnamefont {A.}~\bibnamefont {Lahee}}\
  (\bibinfo  {publisher} {Springer-Verlag Berlin Heidelberg},\ \bibinfo {year}
  {2008})\BibitemShut {NoStop}%
\bibitem [{\citenamefont {Kiriushcheva}\ and\ \citenamefont
  {Kuzmin}(1998)}]{Kiriushcheva1998}%
  \BibitemOpen
  \bibfield  {author} {\bibinfo {author} {\bibfnamefont {N.}~\bibnamefont
  {Kiriushcheva}}\ and\ \bibinfo {author} {\bibfnamefont {S.}~\bibnamefont
  {Kuzmin}},\ }\href {\doibase 10.1119/1.18985} {\bibfield  {journal} {\bibinfo
   {journal} {American Journal of Physics}\ }\textbf {\bibinfo {volume} {66}},\
  \bibinfo {pages} {867} (\bibinfo {year} {1998})}\BibitemShut {NoStop}%
\bibitem [{\citenamefont {Szameit}\ \emph {et~al.}(2011)\citenamefont
  {Szameit}, \citenamefont {Dreisow}, \citenamefont {Heinrich}, \citenamefont
  {Nolte},\ and\ \citenamefont {Sukhorukov}}]{Szameit2011}%
  \BibitemOpen
  \bibfield  {author} {\bibinfo {author} {\bibfnamefont {A.}~\bibnamefont
  {Szameit}}, \bibinfo {author} {\bibfnamefont {F.}~\bibnamefont {Dreisow}},
  \bibinfo {author} {\bibfnamefont {M.}~\bibnamefont {Heinrich}}, \bibinfo
  {author} {\bibfnamefont {S.}~\bibnamefont {Nolte}}, \ and\ \bibinfo {author}
  {\bibfnamefont {A.~A.}\ \bibnamefont {Sukhorukov}},\ }\href {\doibase
  10.1103/PhysRevLett.106.193903} {\bibfield  {journal} {\bibinfo  {journal}
  {Phys. Rev. Lett.}\ }\textbf {\bibinfo {volume} {106}},\ \bibinfo {pages}
  {193903} (\bibinfo {year} {2011})}\BibitemShut {NoStop}%
\bibitem [{\citenamefont {Fassbender}\ \emph {et~al.}(2004)\citenamefont
  {Fassbender}, \citenamefont {Ravelosona},\ and\ \citenamefont
  {Samson}}]{Fassbender2004}%
  \BibitemOpen
  \bibfield  {author} {\bibinfo {author} {\bibfnamefont {J.}~\bibnamefont
  {Fassbender}}, \bibinfo {author} {\bibfnamefont {D.}~\bibnamefont
  {Ravelosona}}, \ and\ \bibinfo {author} {\bibfnamefont {Y.}~\bibnamefont
  {Samson}},\ }\href {http://stacks.iop.org/0022-3727/37/i=16/a=R01} {\bibfield
   {journal} {\bibinfo  {journal} {Journal of Physics D: Applied Physics}\
  }\textbf {\bibinfo {volume} {37}},\ \bibinfo {pages} {R179} (\bibinfo {year}
  {2004})}\BibitemShut {NoStop}%
\bibitem [{\citenamefont {Xia}\ \emph {et~al.}(2010)\citenamefont {Xia},
  \citenamefont {Chun}, \citenamefont {Aizawa}, \citenamefont {Yanagisawa},
  \citenamefont {Krishnan}, \citenamefont {Shindo},\ and\ \citenamefont
  {Tonomura}}]{Xia2010}%
  \BibitemOpen
  \bibfield  {author} {\bibinfo {author} {\bibfnamefont {W.~X.}\ \bibnamefont
  {Xia}}, \bibinfo {author} {\bibfnamefont {Y.~S.}\ \bibnamefont {Chun}},
  \bibinfo {author} {\bibfnamefont {S.}~\bibnamefont {Aizawa}}, \bibinfo
  {author} {\bibfnamefont {K.}~\bibnamefont {Yanagisawa}}, \bibinfo {author}
  {\bibfnamefont {K.~M.}\ \bibnamefont {Krishnan}}, \bibinfo {author}
  {\bibfnamefont {D.}~\bibnamefont {Shindo}}, \ and\ \bibinfo {author}
  {\bibfnamefont {A.}~\bibnamefont {Tonomura}},\ }\href {\doibase
  10.1063/1.3524273} {\bibfield  {journal} {\bibinfo  {journal} {Journal of
  Applied Physics}\ }\textbf {\bibinfo {volume} {108}},\ \bibinfo {eid}
  {123919} (\bibinfo {year} {2010})}\BibitemShut {NoStop}%
\end{thebibliography}

\end{document}